\documentclass[11pt]{article}

\usepackage[margin=1in]{geometry}
\usepackage{graphics}
\usepackage{epsfig}
\usepackage{epstopdf}
\usepackage{amsmath}
\usepackage{array}
\usepackage{subcaption}
\begin{document}
\begin{center}
{\LARGE Expansion Law From First Law of Thermodynamics \\[0.2in]}
{ Mahith M, Krishna P B and Titus K Mathew \\
Department of
Physics, Cochin University of Science and Technology, Kochi-22, India \\ 
mahith15071995@gmail.com; krishnapb@cusat.ac.in; titus@cusat.ac.in}
\end{center}

\date{\today}% It is always \today, today,
%\maketitle

\abstract{Padmanabhan in his paper [arxiv: 1206.4916] put forth an intriguing idea arguing that the accelerated expansion of the universse can 
be viewed as the emergence of cosmic space as cosmic time progresses, with the expansion triggered due to the difference in the degrees of freedom on a 
holographic surface and the one in its emerged bulk. We show that the modified expansion law of the universe proposed by Sheykhi in a non-flat Friedmann- 
Robertson-Walker (FRW) universe and by Cai in a flat FRW universe could be derived starting from the first law of thermodynamics, in (n+1) dimensional 
Einstein gravity, Gauss-Bonnet and more general Lovelock gravity theories. We have also derived the modified versions of the expansion law for a flat 
universe due to Yang et al, in the case of Gauss-Bonnet and Lovelock gravities, from the first law of thermodynamics. This approach is unique in the 
sense that all these modified versions of the expansion law is derived from a thermodynamic identity that has the same form regardless of the gravity 
theory and irrespective of whether the universe is flat or non-flat.}

\section{Introduction}

The past few decades were showered with steadily growing evidence for the existence of a deep connection between gravity and horizon thermodynamics and has turned out to be an exciting realm of research. Following the semi-classical approach, it was discovered by Bekenstein and Hawking \cite{ hawking} that black holes behave similarly to a black body, emitting characteristic wavelength known as 
Hawking radiation with a temperature proportional to the surface gravity at its horizon. Evidently,  black holes also possess an entropy and is 
found to be proportional to the surface area of the horizon\cite{bekenstein}. Trailing along this line of work, the four laws of black hole thermodynamics 
were put forth by Carter, Hawking and Bardeen, that turned out to be analogous to the laws of thermodynamics satisfied by  ordinary macroscopic 
systems\cite{bardeen}. This led to the colossal recognition that black holes behave like an ordinary thermodynamic object.
The thermodynamical analysis of the black hole horizon was later extended to cosmological horizon, for instance in the case of de Sitter universe \cite{Gibbons1}, where the horizon was associated with a temperature called Hawking temperature $T = \frac{1}{2\pi l}$ and an entropy $S = \frac{A}{4G}$, where $ A = 4\pi l^{2}$ is the area of the cosmological event horizon with $l$ being its radius. 
Further major step along this line was taken by Jacobson\cite{Jacobson1}, who derived the Einstein's gravity equation by considering the Clausius relation, 
$\delta Q = TdS $ at the horizon together with the equivalence principle, with $\delta Q$ and $T$ referring to the energy flux and Unruh temperature as perceived by an accelerated observer within the horizon. 
As shown in \cite{paddy}, the Clausius relation also comes into play while interpreting gravitational field equations as an entropy balance law, $\delta S_{m} = \delta S_{grav}$ across a null surface. The thermodynamic perspective of the gravitational field equations have also been investigated by the authors in \cite{sumanta1,sumanta2,paddy5} and that of static spacetimes in \cite{paddy4}.  Moreover, in reference \cite{akbar}, the authors have shown that Friedmann equations at the apparent horizon can be recast to a form given as, $dE=TdS+W dV$, often dubbed as the unified first law of thermodynamics of the universe, where $E$ is the energy of matter within the horizon, $T$ the temperature associated with the horizon, $S$ the associated entropy and $W=(\rho-P)/2,$ the work density with $\rho$ and $P$ referring to the energy density and pressure of matter in the universe.
All the above revelations provokes one to arrive at the astonishing conclusion that gravity is intimately connected with thermodynamics. But, thermodynamics is a macroscopic theory with its variables like pressure, temperature, etc. having no significance in the microscopic realm. These variables which define the macroscopic state of a system "emerges" as a result of the collective behaviour of the constituent microscopic degrees of freedom associated with the system. On similar grounds gravity, as described by Einstein's relativity theory is a macroscopic phenomenon, with its variables including metric, curvature, etc. having no relevance in going over to the microscopic domain. The deep connection between gravity and thermodynamics as mentioned above, then motivates one to reconsider gravity and argue it to be an emergent phenomenon in the same way as thermodynamics.
These insights along with string theory considerations led Verlinde to propose an emergent model of gravity where he interprets it to be an entropic force caused by the changes in the 
information associated with the positions of material bodies\cite{verlinde}. Continuing along this direction, he was able to derive Newton's gravitation law.  A similar line of work was also done by 
Padmanabhan, in which he derived Newton's law of gravity using equipartition law of energy for the degrees of freedom associated with the horizon and the thermodynamic relation 
$S = \frac{E}{2T}$, with $S$ and  $T$ playing the roles of thermodynamic entropy and temperature of the horizon and E being the gravitational mass\cite{paddy2, paddy3}.

Recently Padmanabhan took a step further by arguing cosmic space to be emergent as cosmic time progresses. He asserts that it is conceptually difficult to consider time to be emergent from any pre-geometric variables. This problem can be sorted out in the context of cosmology due to the existence of proper time for the geodesic observers in whose frame CMBR appears to be homogeneous and isotropic\cite{paddy}. He proposes that the accelerated expansion of the universe can be viewed as the emergence of space with the progression of cosmic time.
Padmanabhan put forth such an idea using a specific version of the holographic principle, wherein he had shown that a pure de Sitter universe obeys the holographic equipartition condition of the form $N_{sur} = N_{bulk}$, where $N_{sur}$ is the number of degrees of freedom on the boundary of the Hubble sphere and $N_{bulk}$ is that residing within the boundary. Since the present cosmological observations indicate that our universe is approaching a de Sitter epoch \cite{copeland}, he conjectured that the universe is actually trying to attain this equipartition condition, thereby interpreting the accelerated expansion of the universe being due to the difference in the surface degrees of freedom and the bulk degrees of freedom. Based on this intriguing paradigm, Padmanabhan derived the Friedmann equation of a flat FRW universe in (3+1) Einstein gravity \cite{paddy1}. This idea of Padmanabhan was extended by Cai \cite{cai} to higher dimensional gravity theories like (n+1) dimensional Einstein gravity, Gauss-Bonnet gravity and more general Lovelock gravity for a flat FRW universe by appropriately modifying the surface degrees of freedom on the boundary surface. An extension of this procedure to non-flat FRW universe was done by Sheykhi\cite{sheykhi}. Inspired by Cai's work, an alternative generalization of the dynamical equation due to Padmanabhan was proposed in reference \cite{yang}. More investigations along this line employing Padmanabhan's idea of emergent paradigm can be found in references \cite{sumanta3, sheykhi1, fei, komatsu, tu, fang}. In reference \cite{krishna}, the authors have proved the equivalence between holographic equipartition and entropy maximization. The application of first law of thermodynamics and the validity of generalized second law of thermodynamics have been discussed in \cite{saridakis1,sarkadis2}.
An attempt of deriving the expansion law from the thermodynamic principles was done in reference\cite{dzaki} where the authors used an approximate form of the unified first law of thermodynamics (i.e. $dE=TdS$) as a convenient means of obtaining the expansion law due to Sheykhi in (3+1) Einstein's gravity but failed to extend the idea to higher dimensional gravity theories like the Gauss-Bonnet and Lovelock gravities.

In this paper, starting from the unified first law of thermodynamics $dE=TdS+WdV$ we arrive at Sheykhi's modified version of Padmanabhan's original proposal \cite{sheykhi} in (n+1) dimensional Einstein gravity. We further extend our study to higher order gravity theories like Gauss-Bonnet and Lovelock gravity and derive the corresponding modified expansion laws due to Sheykhi in these gravity theories. Further, we also derive the modified version of the expansion law of a flat FRW universe proposed by Yang et.al \cite{yang} from first law and show that the modifications brought about in the expansion law due to Cai\cite{cai} and Yang et al. \cite{yang} corresponds to each other intimately. We would like to strongly emphasize the fact that even though the expansion law gets modified and generalized to take different forms in different gravity theories for both flat and non-flat universe as shown in references \cite{sheykhi,cai,yang}, the basic thermodynamic relation $dE=TdS+W dV$ retains its form in any gravity theory and irrespective of whether the universe is flat or not.
In the upcoming session, we derive the expansion law due to Sheykhi in (n+1) dimensional Einstein gravity from the first law of thermodynamics. In session 3, we extend our procedure to higher order gravity theories and also show that the modified expansion laws due to Cai\cite{cai} and Yang et al.\cite{yang} corresponds to each other. Finally, we conclude our results in session 4.
\section{Expansion law in (n+1) Einstein gravity from first law of thermodynamics.}
Let us begin by reviewing the basic concepts regarding Padmanabhan's emergent space paradigm \cite{paddy1}. Recent observations show that our universe is asymptotically de Sitter and it satisfies the holographic equipartition condition of the form $N_{sur}=N_{bulk}$ in the final stage. Motivated by this, Padmanabhan argued that the accelerated expansion of the universe is driven by the difference between the surface degrees of freedom on the Hubble horizon and the degrees of freedom in the bulk within the horizon. He therefore proposed the evolution equation of the universe, in (3+1) dimensional Einstein gravity to take the simple form,
\begin{equation}\label{eqn:Law1}
\frac{dV}{dt} = l_{p}^{2} ( N_{sur}-N_{bulk} )
\end{equation}
where $dV$ is the increase in the Hubble volume in an infinitesimal time interval $dt$ and $l_{p}$ is the Planck length.
Here surface degrees of freedom of the Hubble sphere of radius $H^{-1}$ assumes the form,
\begin{equation}
N_{sur} = \frac{4 \pi}{l_{p}^{2}H^{2}} %\cite{paddy}
\end{equation}  
where $l_{p}^{2}$ is the area corresponding to one degree of freedom\\
and the bulk degrees of freedom is given by,
\begin{equation}\label{eqn:bulk}
N_{bulk} =  \frac{|E|}{(1/2)k_{B}T}
\end{equation}
with $|E|=|(\rho+3p)|V$ being the Komar energy and $k_B$ the Boltzmann constant.

Assuming the temperature to be the Hawking temperature $T=H/2\pi$, equation (\ref{eqn:Law1}), along with the continuity equation, $\dot{\rho}+3H(\rho+p)=0$, reduces to the Friedmann equation in (3+1) Einstein gravity.

Cai in \cite{cai} took the first step in extending Padmanabhan's proposal (\ref{eqn:Law1}) to (n+1) dimensional Einstein gravity, Gauss-Bonnet and more general Lovelock gravity theories, by modifying the surface degrees of freedom and the volume increase of the emerged space. He was only able to derive Friedmann equation of a spatially flat FRW universe in these gravity theories.
A successful extension of this procedure for a universe with any spatial curvature was done by Sheykhi in \cite{sheykhi}. He modified Padmanabhan's proposal in (3+1) Einstein gravity (\ref{eqn:Law1}) as,
\begin{equation}\label{eqn:sheykhi}
\frac{dV}{dt}=l_{p}^{2}\frac{r_{A}}{H^{-1}}( N_{sur}-N_{bulk} )
\end{equation}
where
\begin{equation}\label{eqn:apparent}
r_{A}=\frac{1}{\sqrt{H^{2}+\frac{k}{a^{2}}}}
\end{equation}  
is the radius of the apparent horizon, $V=4\pi r_{A}^{3}/3$ is the cosmic volume and $N_{sur}, N_{bulk}$ are defined to be,
\begin{equation}
N_{sur}=\frac{4\pi r_{A}^{2}}{l_{p}^{2}}
\end{equation}
\begin{equation}
N_{bulk}=-\frac{16 \pi^{2}}{3}(\rho+3p)r_{A}^{4}
\end{equation} \\
with $\rho+3p<0$ so as to have $N_{bulk}>0$ \cite{sheykhi},
using which he obtained the Friedmann equation,
\begin{equation}
H^{2}+\frac{k}{a^{2}} = \frac{8\pi l_{p}^{2}}{3}\rho
\end{equation} 
where $k$ here is the spatial curvature of the universe. 
In (n+1) dimensional Einstein gravity he brings a modification to the equation(\ref{eqn:sheykhi}) as,
\begin{equation}\label{eqn:sheykhi1}
\alpha\frac{dV}{dt}=l_{p}^{2}\frac{r_{A}}{H^{-1}}( N_{sur}-N_{bulk} )
\end{equation}
where $ \alpha=(n-1)/2(n-2)$, $V=\Omega_{n}r_{A}^{n}$ is the volume of n-sphere and $N_{sur}, N_{bulk}$ are defined to be,
\begin{equation}
N_{sur}=\alpha \frac{n\Omega_{n}r_{A}^{n-1}}{l_{p}^{2}}
\end{equation}
with $\Omega_{n}$ being the volume of a unit sphere in n-dimensions and,
\begin{equation}\label{eqn:bulk1}
N_{bulk}=-4\pi \Omega_{n}r_{A}^{n+1}\frac{(n-2)\rho+np}{n-2}
\end{equation}
where $(n-2)\rho+np<0$ so as to have $N_{bulk}>0$ \cite{sheykhi}.
\\
Using equation (\ref{eqn:sheykhi1}) he obtained the Friedmann equation,
\begin{equation}
H^{2}+\frac{k}{a^{2}} = \frac{16\pi l_{p}^{n-1}}{n(n-1)}\rho
\end{equation} of (n+1) dimensional FRW universe with any spatial curvature in Einstein gravity.
For Gauss-Bonnet and more general Lovelock gravity theories where entropy does not follow the usual area law, he further brings out a modification to equation (\ref{eqn:sheykhi1}) viz.,
\begin{equation}\label{eqn:sheykhi2}
\alpha\frac{d\tilde{V}}{dt}=l_{p}^{2}\frac{r_{A}}{H^{-1}}( N_{sur}-N_{bulk} )
\end{equation}
where $\tilde{V}$ is the effective volume within the horizon corresponding to the effective area arising from the entropy relation in the gravity theories, and $N_{sur}$ is defined to be \cite{sheykhi},
\begin{equation}
N_{sur}=\frac{\alpha n\Omega_{n}r_{A}^{n+1}}{l_{p}^{n-1}}(r_{A}^{-2}+\alpha r_{A}^{-4}) 
\end{equation}
in case of Gauss-Bonnet gravity and as,
\begin{equation}
N_{sur}=\frac{\alpha n\Omega_{n}r_{A}^{n+1}}{l_{p}^{n-1}} \sum_{i=1}^{m}\hat{c_{i}}r_{A}^{-2i}
\end{equation}
in case of Lovelock gravity,
with $N_{bulk}$ still assuming the form given by equation (\ref{eqn:bulk1}) in both the gravity theories.
Using equation (\ref{eqn:sheykhi2}) he obtained the Friedmann equations of the universe with any spatial curvature in Gauss-Bonnet gravity,
\begin{equation}\label{eqn:sheykhi4}
H^{2}+\frac{k}{a^{2}}+\tilde{\alpha}\Bigg(H^{2}+\frac{k}{a^{2}}\Bigg)^{2} = \frac{16\pi l_{p}^{n-1}}{n(n-1)}\rho
\end{equation}
 and that in the Lovelock gravity,
\begin{equation}\label{equ:sheykhi3}
\sum^{m}_{i=1}\hat{c_{i}}\Big(H^{2}+\frac{k}{a^{2}}\Big)^{i}=\frac{16\pi l_{p}^{n-1}}{n(n-1)} \rho
\end{equation}.
In reference \cite{akbar}, the authors have expressed the Friedmann equation in (n+1) Einstein's gravity, Gauss-Bonnet and more general Lovelock gravity theories as a thermodynamic identity,

\begin{equation}
dE =  TdS + WdV,
\end{equation}
at the apparent horizon of the universe. In the above equation,  $E=\rho V$ is the total energy of matter inside the horizon, $V$ the volume inside the horizon, $W =(\rho - p)/2$ is the work density term \cite{akbar}. This law reduces to the conventional law  $dE=TdS-pdV$ for a pure de Sitter space where $\rho = -p$.
 \\
 
In this section, we will show that the expansion law (\ref{eqn:sheykhi1}) in (n+1) dimensional spacetime due to Sheykhi can be derived starting from the basic thermodynamic relation,
\begin{equation}\label{eqn:TDS}
dE =  TdS + WdV
\end{equation}
For an (n+1) dimensional spacetime, the energy of matter contained in a volume $V=\Omega_{n}r_{A}^{n}$ within the apparent horizon is \cite{akbar},
\begin{equation}\label{eqn:e1}
E = \Omega_{n}r_{A}^{n} \rho 
\end{equation}
where $\Omega_{n}$ is the volume of an n-sphere with unit radius and $r_A$ is the radius of the apparent horizon. Here the entropy of the horizon is proportional to its surface area and is given as \cite{bekenstein},
\begin{equation}
S=\frac{A_h}{4 G}.
\end{equation}
\\Substituting $A_h=n\Omega_n r_A^{n-1}$,$G=l_p^{n-1}$ the entropy of the horizon becomes,
\begin{equation}
S =\frac{n\Omega_{n}r_{A}^{n-1}}{4l_{p}^{n-1}} \hspace{1cm}
\end{equation}
 
The temperature of the horizon is given as \cite{akbar},
\begin{equation}\label{eqn:temp}
T = \frac{\kappa}{2\pi} = \frac{1}{2\pi} \Big[\frac{-1}{r_{A}}\Big(1-\frac{\dot{r_{A}}}{2Hr_{A}}\Big)\Big]
\end{equation}
where $\kappa$ is its surface gravity.

For formulating the thermodynamic identity given by equation (\ref{eqn:TDS}), we obtain $dE$, $TdS$ and $WdV$. Now varying the energy equation (\ref{eqn:e1}), we get $dE$ as,
\begin{equation}
dE=n\Omega_{n}r_{A}^{n-1}\rho dr_{A} + \Omega_{n} r_A^n d\rho
\end{equation}
Using the conservation equation for matter in  $(n+1)$  dimensional spacetime,   $\dot \rho + nH (\rho + p)=0$, the  above equation becomes,
\begin{equation}
dE = n\Omega_{n}r_{A}^{n-1}\rho dr_{A} - n\Omega_{n}r_{A}^{n}(\rho + p )Hdt.
\end{equation}
The work density part $W dV$ in the thermodynamic equation  can be expressed as\cite{akbar},
\begin{equation}
WdV = \Big(\frac{\rho - p}{2}\Big)n \Omega_{n}r_{A}^{n-1}dr_{A}.
\end{equation} 
and the product $TdS$ is given as,
\begin{equation}
\frac{1}{2\pi} \Bigg[\frac{-1}{r_{A}}\Big(1-\frac{\dot{r_{A}}}{2Hr_{A}}\Big)\Bigg]\frac{n(n-1)\Omega_{n}r_{A}^{n-2}dr_{A}}{4l_{p}^{n-1}} 
\end{equation}
Using the above equations the thermodynamic relation (\ref{eqn:TDS}) takes the form,
\begin{equation}\label{eqn:3.28}
\begin{split}
n\Omega_{n}r_{A}^{n-1}\rho dr_{A} - n\Omega_{n}r_{A}^{n}(\rho+p )Hdt = \dfrac{1}{2\pi} \Bigg[\dfrac{-1}{r_{A}}\Big(1-\dfrac{\dot{r_{A}}}{2Hr_{A}}\Big)\Bigg]\dfrac{n(n-1)\Omega_{n}r_{A}^{n-2}dr_{A}}{4l_{p}^{n-1}} 
\\ + \Big(\dfrac{\rho-p}{2}\Big)n \Omega_{n}r_{A}^{n-1}dr_{A}.
\end{split}
\end{equation}\\
Further simplification using equation (\ref{eqn:apparent}) leads to,
\begin{equation}\label{eqn:tds2}
n\Omega_{n}r_{A}^{n+1}\dfrac{8\pi l_{p}^{n-1}(\rho + p)H}{n-1} = n\Omega_{n}r_{A}^{n+1}\Bigg(\dot{H}-\frac{k}{a^{2}}\Bigg)
\end{equation}
( Note that we retain the term $n\Omega_{n}r_{A}^{n+1}$  without canceling since otherwise we have to introduce the same factor by hand to arrive at the result, as can be easily understood in due course.)
Splitting the term on L.H.S and then substituting for the Friedmann equation in (n+1) Einstein gravity 
we get,
\begin{equation}\label{eqn:last}
-n\Omega_{n}r_{A}^{n+1}{\dot{H}} = n\Omega_{n}r_{A}^{n+1}l_{p}^{n-1} \Bigg[\dfrac{H^{2}}{l_{p}^{n-1}} + \dfrac{8\pi[(n-2)\rho + np]}{n(n-1)}\Bigg]
\end{equation}
Substituting for $\dot{H}$ and $H^{2}$ using equation (\ref{eqn:apparent}) we end up in,
\begin{equation}
 n\Omega_{n}\dot{r_{A}}r_{A}^{n-1} = l_{p}^{n-1}\Bigg[\dfrac{ n\Omega_{n}r_{A}^{n-1}r_{A}H}{l_{p}^{n-1}} + \dfrac{8\pi[(n-2)\rho + np]}{n-1}\Omega_{n}r_{A}^{n+2}H \Bigg]
\end{equation}
Using $V=\Omega_{n}r_{A}^{n}$ and $A=n\Omega_{n}r_{A}^{n-1}$ we get,
\begin{equation}
\frac{dV}{dt} =  l_{p}^{n-1}r_{A}H\Bigg[\frac{A}{l_{p}^{n-1}} + \frac{8\pi[(n-2)\rho + np]}{n-1}Vr_{A} \Bigg]
\end{equation}
Looking at the form of the above equation, if us define ,
\begin{equation}
N_{sur} = \frac{\alpha A}{l_{p}^{n-1}} 
\end{equation}
and,
\begin{equation}
N_{bulk} = -\frac{4\pi[(n-2)\rho + np]}{n-2}Vr_{A}
\end{equation}
then we have,

\begin{equation} 
\alpha \frac{dV}{dt} =  l_{p}^{n-1}r_{A}H\Big[N_{sur} - N_{bulk}\Big]
\end{equation}
which is exactly the relation proposed by Sheykhi in \cite{sheykhi} to derive the Friedmann equation in Einstein gravity at the apparent horizon of a non-flat FRW universe.
We would like to highlight the fact that the definitions for $N_{sur}$ and $N_{bulk}$, which were proposed by Sheykhi in his paper \cite{sheykhi}, naturally assumed the correct forms in our derivation, thus enhancing the robustness of our approach.\\
 Also note that if we had considered the case of a flat FRW universe (i.e. $k=0$), then by using equation (\ref{eqn:apparent}) to substitute $r_{A}=H^{-1}$  in equation (\ref{eqn:3.28}) and then following a similar procedure, we would easily obtain,
\begin{equation}\label{eqn:Cailaw}
\alpha \dfrac{dV}{dt} = l_{p}^{n-1}\Big(N_{sur}-N_{bulk} \Big)
\end{equation}\\
which is the expansion law in (n+1) Einstein's gravity as postulated by Cai in \cite{cai}, using which he arrived at the Friedmann equation of a flat FRW universe in (n+1) dimensional Einstein's gravity.
\section{Expansion Law in Gauss-Bonnet  and Lovelock gravity from first law of thermodynamics}
In the previous session we derived the modified expansion law due to Sheykhi from the first law of the thermodynamics in (n+1) Einstein gravity. In this session, we extend the above procedure to other higher dimensional theories like the Gauss-Bonnet and the Lovelock gravity. 
 Gauss-Bonnet gravity, also referred to as Einstein-Gauss-Bonnet gravity, is obtained through the modification of the Einstein-Hilbert action by including the Gauss-Bonnet term \cite{akbar},
 \begin{equation}
 R_{GB} = R^{2} -4R^{\mu \nu}R_{\mu \nu} + R^{\mu \nu \rho \sigma}R_{\mu \nu \rho \sigma}.
\end{equation}
 This term is a topological term in four dimensions and is relevant only in (4+1) dimensions or higher. 
Static black hole solutions and their thermodynamics in vacuum Gauss-Bonnet gravity were examined in reference \cite{boulware, cai9}. 

 In reference \cite{cai1}, the authors have assumed the form of black hole horizon entropy to hold for the apparent horizon of FRW universe as well and is given, in Gauss-Bonnet gravity, as,
 \begin{equation}\label{equ:S}
S = \dfrac{A}{4l_{p}^{n-1}}\Bigg(1+\dfrac{n-1}{n-3}\dfrac{2\tilde{\alpha}}{r_{A}^{2}}\Bigg)
\end{equation}
 where $\tilde{\alpha}=(n-2)(n-3)\alpha$ is known as the Gauss-Bonnet coefficient.\\
Assuming $E$,$T$ and $W$ having the same form as previously discussed but with $S$ taking the form given by equation (\ref{equ:S}), the first law of thermodynamics (\ref{eqn:TDS}) will now become,
\begin{equation}
n\Omega_{n}r_{A}^{n+1}\dfrac{8\pi  l_{p}^{n-1}(\rho + p)}{n-1} =-n\Omega_{n}r_{A}^{n+1}\Bigg(\dot{H}-\frac{k}{a^{2}}\Bigg)\Bigg(1+2\tilde{\alpha}r_{A}^{-2}\Bigg)
\end{equation}

Using Friedmann equation (\ref{eqn:sheykhi4}) in Gauss-Bonnet gravity the above equation can be recast to the form,
\begin{multline}\label{eqn:44}
-n\Omega_{n}r_{A}^{n+1}\dfrac{8\pi  l_{p}^{n-1}((n-2)\rho + np)}{n(n-1)} = n\Omega_{n}r_{A}^{n+1}\dot{H}(1 + 2\tilde{\alpha}r_{A}^{-2}) - \dfrac{k}{a^{2}}\Bigg(n\Omega_{n}r_{A}^{n+1}\Bigg)\Bigg(1 + 2\tilde{\alpha}r_{A}^{-2}\Bigg) + \\ n\Omega_{n}r_{A}^{n+1}\Big(r_{A}^{-2} + \tilde{\alpha}r_{A}^{-4}\Big)
\end{multline}
Substituting for $\dot{H}$ using equation (\ref{eqn:apparent}) the above equation can be put to a suitable form,
\begin{equation}\label{eqn:40}
 n\Omega_{n}r_{A}^{n-1}\dot{r_{A}}(1+2\tilde{\alpha}r_{A}^{-2}) = l_{p}^{n-1}r_{A}H\Bigg[\dfrac{ n\Omega_{n}r_{A}^{n+1} (r_{A}^{-2} +\tilde{\alpha} r_{A}^{-4} )} {l_{p}^{n-1}} + \dfrac{8\pi ((n-2)\rho + np)\Omega_{n}r_{A}^{n+1}}{n-1}\Bigg]
\end{equation}
Looking at the above form of the equation let us define,
\begin{equation}
 N_{sur} =\dfrac{\alpha n\Omega_{n}r_{A}^{n+1}(r_{A}^{-2} +\tilde{\alpha} r_{A}^{-4}) }{l_{p}^{n-1}}
\end{equation}\\
which is identical to the $N_{sur}$ defined by Sheykhi in \cite{sheykhi}
and,
\begin{equation}
 N_{bulk}=-\dfrac{4\pi ((n-2)\rho + np)Vr_{A}}{n-2} 
\end{equation}
where $(n-2)\rho + np <0$ so that $N_{bulk} >0$.\\
Thus equation (\ref{eqn:40}) finally assumes the form,
\begin{equation}\label{eqn:1}
 n\Omega_{n}r_{A}^{n-1}\dot{r_{A}}(1+2\tilde{\alpha}r_{A}^{-2}) = l_{p}^{n-1}r_{A}H\Big(\dfrac{N_{sur}}{\alpha} -\dfrac{N_{bulk}}{\alpha} \Big)
\end{equation}
If we now define an effective area\cite{sheykhi},
\begin{equation}\label{eqn:area}
\tilde{A} =A\Bigg(1+\dfrac{n-1}{n-3}\dfrac{2\tilde{\alpha}}{r_{A}^{2}}\Bigg)
\end{equation}
 for the holographic surface, corresponding to the entropy given by equation (\ref{equ:S}), then the effective volume increase in Gauss-Bonnet gravity is given as,
\begin{equation}\label{eqn:volume}
\dfrac{d\tilde{V}}{dt} = \dfrac{r_{A}}{(n-1)}\dfrac{d\tilde{A}}{dt} = n\Omega_{n}r_{A}^{n-1}\dot{r_{A}}(1+2\tilde{\alpha}r_{A}^{-2})
\end{equation}
which corresponds to the L.H.S of equation (\ref{eqn:1}). Hence, equation (\ref{eqn:1}) can now be conveniently expressed as,
\begin{equation}\label{eqn:sh}
\boxed{\alpha \dfrac{d\tilde{V}}{dt} = l_{p}^{n-1}r_{A}H\Big(N_{sur} - N_{bulk}\Big)}
\end{equation}
which is the proposed expansion law introduced by Sheykhi in order to derive Friedmann equation of a non-flat FRW universe in Gauss-Bonnet gravity at the apparent horizon.
The above discussion again portrays the robustness of our derivation wherein the forms for $N_{surf}$ and $N_{bulk}$ came around naturally in our derivation defining which led us to the correct expansion law of the associated gravity theory.

Again note that if we had considered the case of a flat FRW universe (i.e. $k=0$), then by using equation (\ref{eqn:apparent}) to substitute $r_{A}=H^{-1}$  in equation (3.4) and then following the similar steps, we would easily obtain,
 \begin{equation}\label{eqn:Cailaw2}
\boxed{\alpha \dfrac{d\tilde{V}}{dt} = l_{p}^{n-1}(N_{sur} - N_{bulk})}
 \end{equation}
 which is the equation used by Cai in reference (\cite{cai}) in order to derive the Friedmann equation,
 \begin{equation}\label{eqn:GaussFried}
 H^2+\tilde{\alpha} H^4=\frac{16\pi l_p^{n-1}}{n(n-1)}\rho,
\end{equation} 
 of a spatially flat FRW universe at the Hubble horizon in case of Gauss-Bonnet gravity.

  Inspired by Cai's work, Yang et al. further generalizes the basic dynamical equation (\ref{eqn:Law1}) of Padmanabhan in an (n+1) dimensional universe as  \cite{yang},
\begin{equation}
\dfrac{dV}{dt} = l_{P}^{n-1} f(\Delta N, N_{sur})
\end{equation}
In case of (n+1) dimensional Einstein gravity, the function $f(\Delta N, N_{sur})$ takes the simple form \cite{yang},
\begin{equation}
f(\Delta N) = \Delta N/ \alpha
\end{equation}
with $\Delta N = N_{sur} - N_{bulk}$ and hence give the same relation as equation(\ref{eqn:Cailaw}).
However, in case of Gauss-Bonnet gravity, the function $f(\Delta N, N_{sur})$ takes a complicated form \cite{yang},
\begin{equation}\label{eqn:Yang1}
\boxed{\dfrac{dV}{dt} = l_{p}^{n-1} \dfrac{\dfrac{\Delta N}{\alpha} + \tilde{\alpha}K \Bigg(\dfrac{N_{sur}}{\alpha}\Bigg)^{1+\dfrac{2}{1-n}}}{1+2\tilde{\alpha}K \Bigg(\dfrac{N_{sur}}{\alpha}\Bigg)^{\dfrac{2}{1-n}}}}
\end{equation}
Using the above equation, the authors of \cite{yang} also derived the standard Friedmann equation (\ref{eqn:GaussFried}) of a spatially flat FRW universe in Gauss-Bonnet gravity.

We will now show that equation (\ref{eqn:Yang1}) which corresponds to Yang et al.'s version of the expansion law, can also be derived from the more fundamental thermodynamic identity,
\begin{equation}\label{eqn:de}
 dE =  TdS + WdV.
\end{equation}
Following the procedure carried above, we get from equation (\ref{eqn:44}) (for the flat case),
%\begin{equation}
%n\Omega_{n}r_{A}^{n+1} \Big[\dot{H}(1 + 2\tilde{\alpha}H^{2}) + H^{2}(1 + \tilde{\alpha}H^{2})\Big] = -n\Omega_{n}r_{A}^{n+1}\dfrac{8\pi  l_{p}^{n-1}((n-2)\rho + np)}{n(n-1)}
%\end{equation}
%\flushleft{Thus the thermodynamic relation $dE=TdS+WdV$ can finally be expressed as,}
\begin{equation}\label{eqn:tds4}
n\Omega_{n}r_{A}^{n+1}\Big[-\dot{H}(1+2\tilde{\alpha}H^{2})\Big] =n\Omega_{n}r_{A}^{n+1}\Big[ H^{2} + \dfrac{8\pi  l_{p}^{n-1}((n-2)\rho + np)}{n(n-1)} + \tilde{\alpha}H^{4}\Big]
\end{equation}

which can be further simplified through the steps,
\begin{equation}
-\dot{H}(1+2\tilde{\alpha}H^{2})n\Omega_{n}r_{A}^{n+1} = l_{p}^{n-1}\Bigg[\dfrac{n\Omega_{n}r_{A}^{n-1}}{l_{p}^{n-1}} + \dfrac{8\pi((n-2)\rho + np)}{(n-1)}\Omega_{n}r_{A}^{n+1} + \dfrac{\tilde{\alpha}n\Omega_{n}r_{A}^{n-3}}{l_{p}^{n-1}}\Bigg]
\end{equation}
and
\begin{equation}
-n\Omega_{n}\dot{H}r_{A}^{n+1} = l_{p}^{n-1}\dfrac{\Bigg[\dfrac{n\Omega_{n}r_{A}^{n-1}}{l_{p}^{n-1}} + \dfrac{8\pi((n-2)\rho + np)}{(n-1)}\Omega_{n}r_{A}^{n+1} + \dfrac{\tilde{\alpha}n\Omega_{n}r_{A}^{n-3}}{l_{p}^{n-1}}\Bigg]}{1+2\tilde{\alpha}H^{2}},
\end{equation}
which after some simple rearrangements take the form,
\begin{equation}\label{eqn:76}
-n\Omega_{n}\dot{H}r_{A}^{n+1} = l_{p}^{n-1} \dfrac{\dfrac{n \Omega_{n}r_{A}^{n-1}}{l_{p}^{n-1}}  +  \dfrac{8\pi  \Omega_{n} r_{A}^{n+1}[(n-2)\rho +np]}{n-1} + \bar{\alpha
 }{\Bigg(\dfrac{n\Omega_{n}}{l_{p}^{n-1}})\Bigg)^{\dfrac{2}{n-1}}\Bigg(\dfrac{n \Omega_{n}r_{A}^{n-1}}{l_{p}^{n-1}}\Bigg) ^{1 + \dfrac{2}{1-n}}}}{1+ 2\bar{\alpha
 }{\Bigg(\dfrac{n\Omega_{n}}{l_{p}^{n-1}})\Bigg)^{\dfrac{2}{n-1}}\Bigg(\dfrac{n \Omega_{n}r_{A}^{n-1}}{l_{p}^{n-1}}\Bigg) ^{\dfrac{2}{1-n}}}}
 \end{equation}
where we have substituted for $H$ by $r_{A}^{-1}$ appropriately in the above steps.\\
Note that L.H.S corresponds to $\dfrac{dV}{dt}$ where $V=\Omega_{n}r_{A}^{n}$ and $r_{A}$ = $H^{-1}$ has been used.\\
Now from the above form of the equation, identifying both the degrees of freedom on the horizon surface and the bulk as,
\begin{equation}\label{eqn:nsur12}
N_{sur} = \dfrac{\alpha\hspace{1mm} n\Omega_{n}r_{A}^{n-1}}{l_{p}^{n-1}} = \dfrac{\alpha A}{l_{p}^{n-1}} 
\end{equation}
\begin{equation}
N_{bulk}  = -\dfrac{4\pi V r_{A}[(n-2)\rho + np]}{n-2}
\end{equation}
which can be identified to be the same as in \cite{yang}\\
and also if we let,
\begin{equation}
\Bigg({\dfrac{n\Omega_{n}}{l_{p}^{n-1}}\Bigg)}^{\dfrac{2}{n-1}} = K 
\end{equation}  
equation (\ref{eqn:76}) takes the form,
\begin{equation}\label{eqn:compact}
\dfrac{dV}{dt} = l_{p}^{n-1} \dfrac{\dfrac{N_{sur} - N_{bulk}}{\alpha} + \tilde{\alpha}K \Bigg(\dfrac{N_{sur}}{\alpha}\Bigg)^{1+\dfrac{2}{1-n}}}{1+2\tilde{\alpha}K \Bigg(\dfrac{N_{sur}}{\alpha}\Bigg)^{\dfrac{2}{1-n}}}
\end{equation}
or to a more compact form as,
\begin{equation}\label{eqn:qwerty}
\dfrac{dV}{dt} = l_{p}^{n-1} \dfrac{\dfrac{\Delta N}{\alpha} + \tilde{\alpha}K \Bigg(\dfrac{N_{sur}}{\alpha}\Bigg)^{1+\dfrac{2}{1-n}}}{1+2\tilde{\alpha}K \Bigg(\dfrac{N_{sur}}{\alpha}\Bigg)^{\dfrac{2}{1-n}}}
\end{equation}
with $\Delta N=N_{sur}-N_{bulk}$.
This is identical with the modified Padmanabhan's principle as in reference \cite{yang}. Thus, we have shown that the modified expansion law due to Yang et.al follows from first law of thermodynamics.

We will now show that a mere restructuring of  Yang et al.'s relation (\ref{eqn:qwerty}) followed by a re-definition of the surface degrees of freedom and the volume increase, can result in Cai's version of the modified expansion law (\ref{eqn:Cailaw2}).
For that, we will begin with relation (\ref{eqn:compact}) and substitute the above defined forms of $N_{sur}$ and $K$ for the second term in the numerator. Further, we plug in the forms of $N_{sur}$ and $K$ in the denominator, which is then taken to the L.H.S. to get, after some simplifications,

\begin{equation}
(1+2\tilde{\alpha}H^{2})\dfrac{dV}{dt} = l_{p}^{n-1}\Bigg[ \dfrac{N_{sur} - N_{bulk}}{\alpha} + \tilde{\alpha}\dfrac{\alpha n\Omega_{n}r_{A}^{n-3}}{\alpha l_{p}^{n-1}}\Bigg]
\end{equation}
Using equation (\ref{eqn:nsur12}) we get,
\begin{equation}\label{eqn:gh}
(1+2\tilde{\alpha}H^{2})\dfrac{dV}{dt} = l_{p}^{n-1}\Bigg[ \dfrac{N_{sur} - N_{bulk}}{\alpha} + \tilde{\alpha}\dfrac{r_{A}^{-2} N_{sur} }{\alpha }\Bigg]
\end{equation}
Further using $V = \Omega_{n}r_{A}^{n}$ and $r_{A} = H^{-1}$, the  L.H.S of the above equation reduces to,
\begin{equation}
-n\Omega_{n}r_{A}^{n+1}\dot{H}(1+2\tilde{\alpha}H^{2}) 
\end{equation}
which is equal to $-n\Omega_{n} H^{-n-1}\dot{H}(1+2\tilde{\alpha}H^{2})$.

Eventually, equation (\ref{eqn:gh}) reads,
\begin{equation}\label{eqn:as}
-n\Omega_{n} H^{-n-1}\dot{H}(1+2\tilde{\alpha}H^{2}) = l_{p}^{n-1}\Bigg[ \dfrac{N_{sur} - N_{bulk}}{\alpha} + \tilde{\alpha}\dfrac{r_{A}^{-2} N_{sur} }{\alpha }\Bigg]
\end{equation}
If we now assume the holographic surface to have an effective area $\tilde{A}$ given by equation (\ref{eqn:area}) with $r_{A}$ replaced with $H^{-1}$ instead of the ordinary area A, 
then using the relation between volume V and area A of an n-sphere of radius R given as,
\begin{equation}
\dfrac{dV}{dA} = \dfrac{R}{n-1}
\end{equation} 
one can express the effective volume increase in Gauss-Bonnet gravity using equation (\ref{eqn:volume}), again with $r_{A}$ replaced with $H^{-1}$ as,
\begin{equation}
\dfrac{d\tilde{V}}{dt} = \dfrac{1}{(n-1)H}\dfrac{d\tilde{A}}{dt}
\end{equation}
or,
\begin{equation}
\dfrac{d\tilde{V}}{dt} = \dfrac{-n\Omega_{n}}{H^{n+1}}(1+2\tilde{\alpha}H^{2})\dot{H}
\end{equation}
Subsequently, equation (\ref{eqn:as}) can be written as,
\begin{equation}
\alpha \dfrac{d\tilde{V}}{dt} =l_{p}^{n-1} \Big[N_{sur}(1+\tilde{\alpha}H^{2})-N_{bulk}\Big]
\end{equation}
Substituting for $N_{sur}$ from equation(\ref{eqn:nsur12}) we get:
\begin{equation}
\alpha \dfrac{d\tilde{V}}{dt} = l_{p}^{n-1} \Bigg[\dfrac{\alpha\hspace{1.5mm} n\Omega_{n}r_{A}^{n-1}}{l_{p}^{n-1}} (1+\tilde{\alpha}H^{2})-N_{bulk}\Bigg]
\end{equation}
Looking at the form of the above equation, if we now re-define the surface degrees of freedom associated with the Hubble horizon as,
\begin{equation}
\dfrac{\alpha\hspace{1.5mm} n\Omega_{n}r_{A}^{n-1}}{l_{p}^{n-1}} (1+\tilde{\alpha}H^{2}) = N_{sur} .
\end{equation} 
Then we are finally led to the equation,
\begin{equation}
\alpha \dfrac{d\tilde{V}}{dt} = l_{p}^{n-1} (N_{sur} - N_{bulk})
\end{equation}
which is identical with the modified expansion law (\ref{eqn:Cailaw2}) due to Cai as shown in reference \cite{cai}. \\ \vspace{2mm}

Hence, we conclude that the expansion laws put forth by both Cai and Yang et al. exhibit a strong implicit connection, with the diversity in the two approaches being only due to the difference in the way the authors defined surface degrees of freedom and the rate of volume increase.\vspace{5mm}

Finally, we consider the more general Lovelock gravity theory which generalizes Einstein gravity when spacetime assumes a dimension greater than four. The entropy of a black hole in this theory is shown to have the form \cite{caibh},
\begin{equation}
S=\frac{A_{+}}{4l_{p}^{n-1}}\sum_{i=1}^{m} \frac{i(n-1)}{n-2i+1}\hat{c_{i}}r_{+}^{2-2i}
\end{equation}
where $m=[n/2]$ and the coefficients $\hat{c_{i}}$ are given by,
\begin{equation}
\begin{aligned}\label{eqn:ci}
\hat{c_{0}}=\frac{c_{0}}{n(n-1)}   , \hspace{5mm}  \hat{c_{1}}=1 \\
\hat{c_{i}}=c_{i} \prod_{j=3}^{2m}(n+1-j)\hspace{2mm}   when  \hspace{5mm} i>1
\end{aligned}
\end{equation}
Assuming the black hole entropy formula to hold good for the apparent horizon of the FRW universe as well, with $r_{+}$ replaced by $r_{A}$, then the entropy associated with the apparent horizon is given as,
\begin{equation}\label{entropy}
S=\frac{A}{4l_{p}^{n-1}}\sum_{i=1}^{m} \frac{i(n-1)}{n-2i+1}\hat{c_{i}}r_{A}^{2-2i}
\end{equation}
Extracting expression for effective area $\tilde{A} =4l_{p}^{n-1}S$, from the above form of entropy and using the equation for rate of increase in effective volume \cite{sheykhi},
\begin{equation}
\frac{d\tilde{V}}{dt}=\frac{r_{A}}{(n-1)} \frac{d\tilde{A}}{dt}
\end{equation}
 Sheykhi used the modified expansion law,
\begin{equation}\label{eqn:lovelockexp}
\alpha \dfrac{d\tilde{V}}{dt} = l_{p}^{n-1}r_{A}H(N_{sur} - N_{bulk})
\end{equation}
to arrive at the Friedmann equation,
 \begin{equation}\label{eqn:lovelockFried}
\sum_{i=1}^{m}i\hat{c_{i}}\Bigg(H^{2}+\frac{k}{a^{2}}\Bigg)^{i}=\frac{16\pi l_p^{n-1}}{n(n-1)}\rho,
\end{equation} 
of a universe with any spatial curvature in Lovelock gravity.
%and appropriately defining the number of degrees of freedom on the apparent horizon surface.%

We will now show that equation (\ref{eqn:lovelockexp}) can be obtained from the first law (\ref{eqn:TDS}) in case of Lovelock gravity also. With $E, W$ and $T$ assuming the same form as previously discussed and entropy taking the form given by (\ref{entropy}), equation (\ref{eqn:TDS}) leads us to,
\begin{equation}
-n\Omega_{n}r_{A}^{n+1}\frac{8\pi l_{p}^{n-1}(\rho + p) H}{(n-1)}=n\Omega_{n}r_{A}^{n+1} \Big(\dot{H}-\frac{k}{a^{2}}\Big)\sum_{i=1}^{m}ic_{i}  r_{A}^{2-2i} 
\end{equation}
Substituting using Friedmann equation (\ref{eqn:lovelockFried}) we get, after some simplifications,
%Add $n\Omega_{n}r_{A}^{n+1} \frac{16\pi l_{p}^{n-1} \rho}{n(n-1)}$ on both sides of the above equation which gives,
\begin{equation}\label{eqn:k=0}
-n\Omega_{n}r_{A}^{n+1} \Big(\dot{H}-\frac{k}{a^{2}}\Big) \sum_{i=1}^{m}ic_{i}  r_{A}^{2-2i} = n\Omega_{n}r_{A}^{n+1}l_{p}^{n-1}\Bigg[\sum_{i=1}^{m}\frac{ c_{i}\Big(H^{2}+\frac{k}{a^{2}}\Big)^{i}}{l_{p}^{n-1}} + \frac{8\pi ((n-2)\rho + np) }{n(n-1)}\Bigg]
\end{equation}
Further using equation (\ref{eqn:apparent}) one will finally end up in,
\begin{equation}\label{eqn:llg}
n\Omega_{n}r_{A}^{n+1}\dot{r_{A}} \sum_{i=1}^{m}ic_{i}r_{A}^{-2i}=r_{A}Hl_{p}^{n-1}\Bigg[n\Omega_{n}r_{A}^{n+1} \sum_{i=1}^{m}\frac{ c_{i}r_{A}^{-2i}}{l_{p}^{n-1}} + \frac{\Omega_{n}r_{A}^{n+1} 8\pi ((n-2)\rho + np) }{(n-1)}\Bigg]
\end{equation}
Looking at the above form of the equation, if us define,
\begin{equation}
N_{sur}=\frac{\alpha n\Omega_{n}r_{A}^{n+1}}{l_{p}^{n-1}} \sum_{i=1}^{m}c_{i}r_{A}^{-2i}
\end{equation}
and,
\begin{equation}
N_{bulk} =  -\dfrac{4\pi \Omega_{n}r_{A}^{n+1}[(n-2)\rho + np]}{n-2} 
\end{equation}
the above equation takes the form,
\begin{equation}\label{eqn:llg11}
n\Omega_{n}r_{A}^{n+1}\dot{r_{A}} \sum_{i=1}^{m}ic_{i}r_{A}^{-2i}=l_{p}^{n-1}r_{A}H\Bigg(\frac{N_{sur}}{\alpha}-\frac{N_{bulk}}{\alpha}\Bigg)
\end{equation}
From the entropy relation (\ref{entropy}), the effective area of the apparent horizon is given by\cite{sheykhi},
\begin{equation}
\tilde{A}=n\Omega_{n}r_{A}^{n-1}\sum_{i=1}^{m} \frac{i(n-1)}{n-2i+1}\hat{c_{i}}r_{A}^{2-2i}
\end{equation}
Then the increase in the effective volume of the apparent horizon in Lovelock gravity is,
\begin{equation}
\frac{d\tilde{V}}{dt}=n\Omega_{n}r_{A}^{n+1}\dot{r_{A}} \sum_{i=1}^{m}ic_{i}r_{A}^{-2i}
\end{equation} 
which corresponds to the L.H.S of equation (\ref{eqn:llg11}).\\
Hence equation (\ref{eqn:llg11}) finally become,
\begin{equation}\label{eqn:lovelock}
\alpha\frac{d\tilde{V}}{dt}=l_{p}^{n-1}r_{A}H\Big(N_{sur}-N_{bulk}\Big)
\end{equation}
which is nothing but the expansion law in Lovelock gravity used in reference \cite{sheykhi} in order to derive the Friedmann equation of a non-flat FRW universe in Lovelock gravity theory.
Again for the special case of $k=0$, we get,
\begin{equation}\label{eqn:94}
\alpha\frac{d\tilde{V}}{dt}=l_{p}^{n-1}\Big(N_{sur}-N_{bulk}\Big)
\end{equation}
 which was the equation used by Cai in reference \cite{cai} in order to obtain the Friedmann equation of a flat FRW universe in Lovelock gravity.
 
 Inspired by Cai's work, Yang et.al generalized the basic dynamical equation (\ref{eqn:Law1}) of Padmanabhan in case of a flat FRW universe in Lovelock gravity too, as \cite{yang},
\begin{equation}\label{eqn:95}
\dfrac{dV}{dt} = l_{P}^{n-1} f(\Delta N, N_{sur})
\end{equation}
with $f(\Delta N, N_{sur})$ given by,
\begin{equation}
f(\Delta N, N_{sur})=\frac{\Delta N/\alpha+\sum_{i=2}^{m}\hat{c_{i}}k_{i}(N_{sur}/\alpha)^{1+\frac{2i-2}{1-n}}}{1+\sum_{i=2}^{m}i\hat{c_{i}k_{i}(N_{sur}/\alpha)^{\frac{2i-2}{1-n}}}}
\end{equation} 
where $k_{i} =\Big(n\Omega_{n}/l_{p}^{n-1}\Big)^{\frac{2i-2}{1-n}}$, $m=[n/2]$ and $\hat{c_{i}}$ are some coefficients as mentioned previously in equation (\ref{eqn:ci}).
We will now show that equation (\ref{eqn:95}) readily follows from the first law, $dE=TdS+WdV$, just like in the Gauss-Bonnet case.
Beginning with the first law and following the same steps as before, we end up in equation (\ref{eqn:k=0}) which for the flat case gives,
\begin{equation}\label{eqn:LHS}
-n\Omega_{n}r_{A}^{n-1}\sum_{i=1}^{m}i\hat{c_{i}}\dot{H}H^{2i-4}=l_{p}^{n-1}\Bigg[\frac{n\Omega_{n}r_{A}^{n-1}\sum_{i=1}^{m}\hat{c_{i}}H^{2i-2}}{l_{p}^{n-1}}+\frac{\Omega_{n}r_{A}^{n+1}8\pi[(n-2)\rho+np]}{n-1}\Bigg]
\end{equation}
Using $V=\Omega_{n}r_{A}^{n}$, L.H.S can be simplified to give,
\begin{equation}
\frac{dV}{dt}\Bigg[1+\sum_{i=2}^{m}i\hat{c_{i}}\Bigg(\frac{n\Omega_{n}}{l_{p}^{n-1}}\Bigg)^{\frac{2i-2}{n-1}}\Bigg(\frac{n\Omega_{n}r_{A}^{n-1}}{l_{p}^{n-1}}\Bigg)^{\frac{2i-2}{1-n}}\Bigg]
\end{equation}
Therefore equation (\ref{eqn:LHS}) will become, after some minor simplications,
\begin{equation}
\begin{split}
\frac{dV}{dt}\Bigg[1+\sum_{i=2}^{m}i\hat{c_{i}}\Bigg(\frac{n\Omega_{n}}{l_{p}^{n-1}}\Bigg)^{\frac{2i-2}{n-1}}\Bigg(\frac{n\Omega_{n}r_{A}^{n-1}}{l_{p}^{n-1}}\Bigg)^{\frac{2i-2}{1-n}}\Bigg]=l_{p}^{n-1}\Bigg[\frac{n\Omega_{n}r_{A}^{n-1}}{l_{p}^{n-1}}+\frac{\Omega_{n}r_{A}^{n+1}8\pi[(n-2)\rho+np]}{n-1}+\\ \sum_{i=2}^{m}\hat{c_{i}}\Bigg(\frac{n\Omega_{n}}{l_{p}^{n-1}}\Bigg)^{\frac{2i-2}{n-1}}\Bigg(\frac{n\Omega_{n}r_{A}^{n-1}}{l_{p}^{n-1}}\Bigg)^{1+\frac{2i-2}{1-n}}\Bigg]
\end{split}
\end{equation}
Looking at the above form of the equation, let us define,
\begin{equation}\label{eqn:nsur}
N_{sur}=\frac{\alpha n\Omega_{n}r_{A}^{n-1}}{l_{p}^{n-1}}
\end{equation}
\begin{equation}
 N_{bulk}= -\dfrac{4\pi V r_{A}[(n-2)\rho + np]}{n-2} 
\end{equation}
and
\begin{equation}
 K = \Bigg({\dfrac{n\Omega_{n}}{l_{p}^{n-1}}\Bigg)}^{\dfrac{2i-2}{n-1}} 
\end{equation}  
then we easily recover Yang et al.'s proposed relation,
\begin{equation}\label{eqn:103}
\frac{dV}{dt}=l_{p}^{n-1}\Bigg[\frac{\Delta N/\alpha+\sum_{i=2}^{m}\hat{c_{i}}k_{i}(N_{sur}/\alpha)^{1+\frac{2i-2}{1-n}}}{1+\sum_{i=2}^{m}i\hat c_{i}k_{i}{(N_{sur}/\alpha)}^{\frac{2i-2}{1-n}}}\Bigg]
\end{equation}
This further confirms our main objective that any modifications to Padmanbhan's original conjucture can easily be derived from the first law of thermodynamics with the forms for $N_{sur}$ and $N_{bulk}$ following naturally and elegantly, as shown in our derivation.
Finally, we will again show that Cai's relation (\ref{eqn:94}) of the expansion law in Lovelock gravity follows from Yang et al.'s relation (\ref{eqn:103}). For that we start with equation (\ref{eqn:103}) that can be rewritten as,
\begin{equation}
\frac{dV}{dt}\Bigg[1+\sum_{i=2}^{m}i\hat c_{i}k_{i}{(N_{sur}/\alpha)}^{\frac{2i-2}{1-n}}\Bigg]=l_{p}^{n-1}\Bigg[\frac{(N_{sur}-N_{bulk})}{\alpha}+\sum_{i=2}^{m}\hat{c_{i}}k_{i}{(N_{sur}/\alpha)}^{1+\frac{2i-2}{1-n}}\Bigg]
\end{equation}
Substitute for $K_{i}$ and $N_{sur}$ on L.H.S and in the second term on R.H.S we get,
\begin{equation}\label{eqn:effv}
-\alpha \frac{n\Omega_{n}}{H^{n+3}}\dot{H}\Bigg(\sum_{i=1}^{m}i\hat{c_{i}}H^{2i}\Bigg) = l_{p}^{n-1}\Bigg[N_{sur}\Bigg(\sum_{i=1}^{m}\hat{c_{i}}H^{2i-2}\Bigg)-N_{bulk}\Bigg]
\end{equation}
If we define an effective area for the horizon,
\begin{equation}
\tilde{A}=\frac{n\Omega_{n}}{H^{n-1}}\sum_{i=1}^{m}\frac{i(n-1)}{n-2i+1}\hat{c_{i}}H^{2i-2}
\end{equation}
corresponding to the entropy relation (\ref{entropy}), then the effective volume increase is given as,
\begin{equation}
\frac{d\tilde{V}}{dt}=-\frac{n\Omega_{n}}{H^{n+3}}\dot{H}\Bigg(\sum_{i=1}^{m}i\hat{c_{i}}H^{2i}\Bigg)
\end{equation}
Therefore equation (\ref{eqn:effv}) will now become,
\begin{equation}
\alpha \frac{d\tilde{V}}{dt}=l_{p}^{n-1}\Bigg[N_{sur}\Bigg(\sum_{i=1}^{m}\hat{c_{i}}H^{2i-2}\Bigg)-N_{bulk}\Bigg]
\end{equation}
Substitute for $N_{sur}$ using equation (\ref{eqn:nsur}) we get,
\begin{equation}
\alpha \frac{d\tilde{V}}{dt}=l_{p}^{n-1}\Bigg[\frac{\alpha n\Omega_{n}r_{A}^{n-1}}{l_{p}^{n-1}}\Bigg(\sum_{i=1}^{m}\hat{c_{i}}H^{2i-2}\Bigg)-N_{bulk}\Bigg]
\end{equation}
 Looking at the above form of the equation if we now redefine $N_{sur}=\frac{\alpha n\Omega_{n}r_{A}^{n-1}}{l_{p}^{n-1}}\Bigg(\sum_{i=1}^{m}\hat{c_{i}}H^{2i-2}\Bigg)$, we eventually get,
 \begin{equation}
 \alpha \frac{d\tilde{V}}{dt}=l_{p}^{n-1}(N_{sur}-N_{bulk})
 \end{equation}
 which is Cai's relation \cite{cai}.
 Thus, we conclude that a mere restructuring of Yang's relation along with a re-definition of the surface degrees of freedom and the volume increase will lead to Cai's relation in Lovelock gravity.
 \section{Conclusion}
The novel idea that the accelerated expansion of the universe could be viewed as the emergence of space was first introduced by Padmanabhan in \cite{paddy}, where he showed that the dynamical equation governing the evolution of a flat FRW universe in (3+1) Einstein gravity could be derived from his proposed expansion law. This idea was extended to derive the Friedmann equations of a flat universe in higher dimensional theories like the (n+1) Einstein gravity, Gauss-Bonnet and more general Lovelock gravity theories by Cai. In reference \cite{cai} Cai altered Padmanabhan's version of the expansion law and appropriately modified the volume increase of the emerged space and the surface degrees of freedom. However, instead of modifying the volume increase and the degrees of freedom on the holographic surface, Yang et al. \cite{yang} generalized the expansion law itself by introducing functions that assumes different forms in different gravity theories. Padmanabhan's idea was extended to a non-flat FRW universe by Sheykhi \cite{sheykhi}. By modifying Padmanabhan's proposal he obtained Friedmann equations in the case of (n+1) Einstein, Gauss-Bonnet and  Lovelock gravity theories. Interestingly, the main objective of all the above authors were the same, that is to derive Friedmann equations of appropriate gravity theories by aptly proposing a suitable form for the expansion law instead of deriving them.

 The central theme of our paper is that all these differently modified versions of the expansion law could be derived from the fundamental thermodynamic relation, $dE = TdS + WdV$, for both flat and non-flat universe. In this paper, we have shown that the modified expansion law for a non-flat universe proposed by Sheykhi in the case of (n+1) Einstein gravity, Gauss-Bonnet and more general Lovelock gravity theories can be derived from the thermodynamic identity, $dE = TdS + WdV$. Further we have shown that Cai's modified expansion law for a flat universe follows from the non-flat case once we let the curvature constant, $k$, in the expression for apparent radius to vanish. Yang et al. put forth an alternative form of the expansion law for a flat universe in Gauss-Bonnet and Lovelock gravities, that differed significantly from Cai's expansion law. We also derived this modified version of the expansion law of Yang et al in the case of Gauss-Bonnet and Lovelock gravity theories starting from the thermodynamic relation. It also became clear that the approaches employed  by Cai and Yang et al. only differed in their definition of the rate of volume increase and surface degrees of freedom associated with the horizon of the FRW universe.
 
In summary, there are various forms of the expansion law that give the same Friedmann equations in different gravity theories \cite{cai,yang}. Also, the expansion law takes  different forms for flat and non-flat FRW universe \cite{cai,yang,sheykhi} as well. The underlying factor that highlights the uniqueness of our approach is that the very form of the fundamental thermodynamic relation, $dE = TdS + WdV$, remains the same irrespective of the gravity theory for both flat and non-flat universe. The procedure followed in this paper and the results hence obtained depicts that our approach is so powerful enough that it could be used to derive modified expansion laws of the universe with or without a spatial curvature, in any gravity theory.
% Non-BibTeX users please us


\begin{thebibliography}{}
\bibitem{hawking} S.W Hawking, Commun.Math.Phys.43,199(1975)
	\bibitem {bekenstein} J.D.Bekenstein, Phys.Rev.D7,2333(1973)
	\bibitem{bardeen} J.M.Bardeen, B.Carter and S.W.Hawking, \textit{The four laws of black hole mechanics}, Commun.Math.Phys.31,161(1973).
	\bibitem{Gibbons1} G. W. Gibbons and S.W. Hawking, \textit{Cosmological event horizons, thermodynamics and particle creation}, Phys. Rev. D15, 2738 (1977).
	\bibitem{Jacobson1} Ted Jacobson, \textit{Thermodynamics of Spacetime: The Einstein Equation of State}, Phys. Rev. Lett. 75, 1260 (1995)[arxiv:hep-th/0212327]
	\bibitem{paddy} T.Padmanabhan, \textit{A Physical Interpretation of Gravitational Field Equations}, AIp Conf. proc. 1241, 93 (2010) [arxiv:0911.1403 [gr-qc]]; T.Padmanabhan, \textit{Entropy density of spacetime and thermodynamic interpretation of field equations of gravity in any diffeomorphic invariant theory}, [arxiv:0903.1254[hep-th]]
	\bibitem{sumanta1} Sumanta Chakraborty, Krishnamohan Parattu and T. Padmanabhan, \textit{Gravitational field equations near an arbitrary null surface expressed as a thermodynamic identity}, JHEP 10(2015)097
	\bibitem{sumanta2} Sumanta Chakraborty , \textit{Lanczos-Lovelock gravity from a thermodynamic perspective}, JHEP 1508, 029 (2015)
	\bibitem{paddy5} T.Padmanabhan, Mod. Phys. Lett. A 25 1129 (2010) [arxiv: 0912.3165]; Phys. Rev. D 81 124040 (2010) [arxiv: 1003.5665]
	\bibitem{paddy4} T.Padmanabhan, Class.Quan.Grav., 21, 4485 (2004) [gr-qc/0308070]
	
	\bibitem{akbar} M.Akbar and Rong-Gen Cai, \textit{Thermodynamic behaviour of Friedmann equations at the apparent horizon of FRW Universe}, Phys.Rev.D75:084003, (2007)
	\bibitem{verlinde} E.p. Verlinde, On the origin of gravity and the laws of Newton, JHEP 1104, 029(2011) [arxiv:1001.0785[hep-th]]
	\bibitem{paddy2} T. Padmanabhan, \textit{Equipartition of energy in the horizon degrees of freedom and the emergence of gravity}, Mod. Phys. Lett. A 25, 1129 (2010) [arXiv:0912.3165 [gr-qc]].
	\bibitem{paddy3} T. Padmanabhan, \textit{Gravitational entropy of static spacetimes and microscopic density of states}, Class. Quant. Grav.21, 4485 (2004) [gr-qc/0308070].
	\bibitem{copeland} Edmund J. Copeland, M. Sami, Shinji Tsujikawa, \textit{Dynamics of dark energy}, Int.J.Mod.Phys.D15:1753-1936, (2006)
	
	\bibitem{paddy1} T.Padmanabhan, \textit{Emergence and Expansion of Cosmic Space as due to the Quest for Holographic Equipartition}, arxiv:1206.4916 [hep-th]
	\bibitem{cai} Rong-Gen Cai, \textit{Emergence of Space and Spacetime Dynamics of Friedmann-Robertson-Walker Universe}, JHEP11(2012)016 
	\bibitem{sheykhi} Ahmed Sheykhi, \textit{Friedmann equations from emergence of cosmic space}, Phys. Rev.D 87, 061501(R) (2013)
	\bibitem{yang} Ke Yang, Yu-Xiao Liu, Yong-Qiang Wang, \textit{Emergence of Cosmic Space and the Generalized Holographic Equipartition}, Phys. Rev. D 86, 104013 (2012)
	\bibitem{sumanta3} Sumanta Chakraborty, T. Padmanabhan, \textit{Evolution of Spacetime arises due to the departure from Holographic Equipartition in all Lanczos-Lovelock Theories of Gravity}, Phys. Rev. D 90 124017 (2014).
	\bibitem{sheykhi1} Ahmad Sheykhi, \textit{Modified Friedmann Equations from Tsallis Entropy}, [arXiv:1806.03996v3]
	\bibitem{fei} Fei-Quan Tu, Yi-Xin Chen, Bin Sun, You-Chang Yang, \textit{Accelerated expansion of the universe based on emergence of space and thermodynamics of the horizon}, Physics Letters B, Volume 784, Pages 411-416, (10 September 2018)
	\bibitem{komatsu} Nobuyoshi Komatsu, \textit{Thermodynamic constraints on holographic-principle-based cosmological scenarios: entropic-force models and holographic equipartition models}, [arXiv:1810.11138v1 [gr-qc]]
	\bibitem{tu}  Fei-Quan Tu and Yi-Xin Chen, \textit{Emergence of spaces and the dynamic equations
		of FRW universes in the f(R) theory and deformed Hořava-Lifshitz theory}, JCAP05(2013)024.
	\bibitem{fang} Fang-Fang Yuan and Peng Huang, \textit{Emergent cosmic space in Rastall theory},  [arXiv:1607.04383v3 [gr-qc]]
	\bibitem{krishna} Krishna P B and Titus K Mathew, \textit{Holographic Equipartition and the Maximization of Entropy}, Phys. Rev. D 96.063513 (February 2017)
	\bibitem{saridakis1} Mubasher Jamil, Emmanuel N. Saridakis and M. R. Setare,\textit{Thermodynamics of dark energy interacting with dark matter and radiation}, Phys.Rev. D81 (2010) 023007 [arXiv:0910.0822v3 [hep-th]]
	\bibitem{sarkadis2} Andreas Lymperis and Emmanuel N. Saridakis, \textit{Modified cosmology through nonextensive horizon thermodynamics}, [arXiv:1806.04614v1 [gr-qc]]
	
\bibitem{dzaki} Fatemeh Lalehgani Dezaki and Behrouz Mirza, \textit{Generalized entropies and the expansion law of the universe}, Gen Relativ Gravit (2015) 47:67, [ arXiv:1406.3712v2 [gr-qc]]
	
	\bibitem{cai1} Rong-Gen Cai and Sang Pyo Kim, \textit{ First law of thermodynamics and Friedmann equations of FRW Universe}, JHEP 0502 (2005) 050 hep-th/0501055 

	\bibitem{boulware} D.G Boulware and S. Desre, Phys. Rev. Lett. 55, 2656 (1985); J.T Wheeler, Nucl. Phys. B 268, 737 (1986); Nucl. Phys. B 273, 732 (1986);  R.C Myers and J.Z Simon, Phys. Rev. D 38, 2434 (1988).
	\bibitem{cai9} R.G Cai, Phys. Rev. D65, 084014 (2002); R.G Cai and Q. Guo, Phys. Rev. D69, 104025 (2004)
 
	\bibitem{caibh} R.G Cai, \textit{A note on thermodynamics of black holes in Lovelock gravity}, Phys. Lett. B 582, 237 (2004).
	

\end{thebibliography}
\end{document}